\documentclass[pra,twocolumn,aps]{revtex4}

\usepackage[T1]{fontenc}
\usepackage[latin1]{inputenc}
 \usepackage{bm}
 \usepackage{amsmath}
 \usepackage{amssymb}
 \usepackage{latexsym}
 \usepackage{amsfonts}
 \usepackage{epsfig}
 \usepackage{psfrag}

 \newcommand{\nn}{\nonumber\\}

 \newcommand{\vau}{\mbox{\boldmath$v$}}
 \newcommand{\na}{\mbox{\boldmath$\nabla$}}
 \newcommand{\bea}{\begin{eqnarray}}
 \newcommand{\ea}{\end{eqnarray}}
 \newcommand{\eea}{\end{eqnarray}}
 \newcommand{\ord}{{\cal O}}

\begin{document}
  
\title{Quantum fluctuations in trapped time-dependent
	Bose-Einstein condensates}

\author{Michael Uhlmann}

\affiliation{Department of Physics, Australian National
University, Canberra ACT 0200, Australia}

\begin{abstract}

Quantum fluctuations in time-dependent, harmonically-trapped
Bose-Einstein condensates are studied within Bogoliubov theory.
An eigenmode expansion of the linear field operators permits the
diagonalization of the Bogoliubov-de Gennes equation for a stationary
condensate.
When trap frequency or interaction strength are varied, the
inhomogeneity of the background gives rise to off-diagonal coupling
terms between different modes.
This coupling is negligible for low energies, i.e.,
in the hydrodynamic regime, and an effective space-time metric can be
introduced.
The influence of the inter-mode coupling will be demonstrated in an example,
where I calculate the quasi-particle number for a quasi-one-dimensional
Bose-Einstein condensate subject to an exponential sweep of interaction
strength and trap frequency.

\end{abstract}

  
\maketitle


\section{Introduction}

Ultracold atomic gases offer various opportunities for the study of
interacting many-body quantum systems
%
in a well-controlled environment
\cite{BEC,Manybody}.
For instance, the Bose-Hubbard model -- a simplified description for
bosons in a periodic potential -- can be studied with Bose-Einstein
condensates confined in optical lattices \cite{Mott,BHM}.
%
%
%
%
%
%
Quantum gases
have also gained much attention lately regarding
the emergence of an effective space-time
\cite{Unruh,Garay,Review,Piyush,Silke,Barcelo,PeterF,Uwe,Uwe2,ich_njp,
Kurita,WuesterBlackhole,Matt,Volovik,artificialblackholes}:
their low-energy phase fluctuations
obey the same covariant field equations as a
scalar quantum field in a certain curved space-time.
Hence, the study of phonons in this laboratory system might shed some light
on aspects of cosmic quantum effects, e.g., Hawking radiation
\cite{Hawking,Garay,WuesterBlackhole,artificialblackholes,Unruh}
or the freezing and amplification
of quantum fluctuations in expanding spacetimes
\cite{LL,PeterF,Uwe,ich_njp,Silke,Piyush,Barcelo,BirrellDavis}.
Although the fluctuations in Bose-Einstein condensates are usually small,
it has recently become possible in experiments to go beyond the
classical order parameter and resolve signatures of the fluctuations
\cite{noise-correlations,interference,Bragg,ultracoldHBT,numberfluct}.
%

Theoretically, the fluctuations in a Bose-Einstein condensate are usually
treated as small perturbations of the mean field.
The solution of the coupled field equations 
is rather demanding and often requires further approximations,
especially for time-dependent condensates.
The Hartree-Fock-Bogoliubov method, see, e.g., \cite{Griffin},
permits in principle the self-consistent propagation of the
mean field and the quantum correlations for arbitrary
time-dependences of the trap potential or interaction strength.
But the scaling of the numerics with system size often limits the
actual calculations to a low number of dimensions, certain symmetries, or
a short time interval.
Thermal condensates might be studied using the projected
Gross-Pitaevskii equation \cite{projectedGP},
where the low-energy
part of the fluctuations is expanded into an arbitrary basis and, in
view of the large thermal occupation, treated classically; higher
excitations as well as the vacuum effect are omitted.
On the other hand,  studies in the context of expanding spacetimes, e.g.,
\cite{Silke,Piyush,ich_njp,Uwe,Barcelo,Review,Uwe2},
indeed focus on the quantum fluctuations but often assume a
homogeneous background or start with the hydrodynamic
action, which is only valid on scales longer than the healing length.

In this article, I will discuss the evolution of the quantum fluctuations
in trapped time-dependent Bose-Einstein condensates.
The linear field operators will be expanded into their eigenmodes thus
permitting the diagonalization of the (initial) evolution equations.
Although basis expansions of the field operator are frequently used
(e.g., in \cite{projectedGP,WuesterBosenova}),
these references usually consider the harmonic oscillator eigenfunctions,
a large number of which must be used in order to describe
the excitations properly.
By adopting the eigenmodes, a much smaller part of the basis needs to be
considered and many more situations will become numerically feasible.
(Note, however, that in order to obtain the fluctuation eigenmodes,
a relatively large number of oscillator functions must be employed --
but they need not be propagated.)

This Article is organized as follows.
Section \ref{SECfieldequations} reviews 
the field equations and their linearization.
The Bogoliubov-de Gennes equation for the linear quantum fluctuations
can be diagonalized by an eigenmode expansion, which will be performed
in Sec.\ \ref{SECeigenmode}.
However, as soon as trap frequency or interaction strength are varied,
off-diagonal terms appear.
This coupling of different modes is negligible for excitations
with energies much smaller than the chemical potential even in time-dependent
condensates, as will be shown in Sec.\ \ref{SECTF}, where the order parameter
is treated in the Thomas-Fermi approximation.
It is then also possible to establish the analogy between phase fluctuations
and a massless scalar field in a certain curved spacetime.
If the excitation energies are of the same order as the chemical potential,
the coupling of different modes might lead to a population transfer,
which will be illustrated in an example in Sec.\ \ref{SEC1D}.
%


\section{Field equations}
\label{SECfieldequations}

\subsection{Scaling transformation}

In dimensionless units, the field operator $\hat \Psi$ of a trapped
(quasi)-$D$-dimensional
Bose-Einstein condensate obeys the non-linear Schr\"odinger field equation
\cite{dimless}
    \begin{alignat}{2}
	i \frac{\partial}{\partial t} \hat \Psi \, = \,
		\left[ - \frac{\na^2}{2} + \omega^2(t) \frac{\bm r^2}{2}
			+ g(t) \hat \Psi^\dagger \hat \Psi \right]
		\hat \Psi \,,
	\label{fieldequation}
    \end{alignat}
with $D$-dimensional coupling strength $g(t) \propto a_s$.
By changing the $s$-wave scattering length $a_s$ through Feshbach resonances
\cite{feshbach} or by varying the trap frequency $\omega(t)$, an external
time-dependence can be prescribed on the condensate.
The gas cloud will adapt to these changes and it will either expand or
contract and with it the quasi-particle excitations residing upon it.
A part of this background motion can be accounted for by transforming to
new coordinates $\bm x = \bm r/ b(t)$ with scale factor $b(t)$.
The field operator then reads \cite{scaling} 
    \begin{alignat}{2}
	\hat \Psi(\bm r, t) \, = \,
		e^{i\Phi} \frac{\hat \psi(\bm x,t)}{b^{D/2}} \,.
	\label{scaling}
    \end{alignat}
The phase $\Phi = (\bm r^2/2) \dot b /b$ is chosen such as to generate
an isotropic velocity field $\na \Phi = \bm r\dot b/b$ which (at least
partially) describes the expansion/contraction of the condensate.
If the scale factor $b(t)$ obeys
    \begin{alignat}{2}
	f^2(t) \, = \,
		b^3 \frac{\partial^2 b}{\partial t^2} + b^4 \omega^2(t)
		\, = \, \frac{g(t)}{g_0} b^{2-D} \,,
	\label{fvont}
    \end{alignat}
with $g_0$ being the initial value of the coupling strength, a scaled
field equation follows
    \begin{alignat}{2}
	i b^2 \frac{\partial}{\partial t} \hat \psi \, = \,
		\left[
			- \frac{\na_{\bm x}^2}{2}
			+ f^2 \left( \frac{\bm x^2}{2} +
				g_0 \hat \psi^\dagger \hat \psi\right)
		\right] \hat \psi \,,
	\label{scaledHeisenberg}
    \end{alignat}
where trapping and interaction terms have acquired the same time-dependent
pre-factor $f^2(t)$ and all other coefficients are time-independent.
(The scale factor $b(t)$ on the left hand side might be included into
a redefined time $d\tau = dt/b^2$, see Sec.\ \ref{SECspacetime}.)

\subsection{Linearization}

For large particle numbers $N$, one might formally expand the field operator
into inverse powers of $\sqrt{N}$ \cite{meanfield}
    \begin{alignat}{2}
	\hat \psi \, = \, \left( \psi_0 + \hat \chi + \hat \zeta \right)
		\frac{\hat A}{\sqrt{\hat N}} \,.
	\label{meanfield}
    \end{alignat}
Here, $\hat A$ and $\hat N = \hat A^\dagger \hat A$ are the atomic operators.
They commute with the linear $\hat \chi = \ord(N^0)$ and higher-order quantum
excitations $\hat \zeta$ and thus yield the exact conservation of particle
number.
The order parameter $\psi_0 = \ord(\sqrt N)$ in the center of the trap
but diminishes towards the edge of the condensate.
Insertion of the expansion \eqref{meanfield} into the scaled Heisenberg
equation \eqref{scaledHeisenberg} yields the Gross-Pitaevskii equation
for the classical background $\psi_0$ \cite{GP}
    \begin{alignat}{2}
	i b^2 \frac{\partial}{\partial t} \psi_0 \, = \,
		\left[ - \frac{\bm \nabla_x^2}{2}
			+ f^2 \left( \frac{\bm x^2}{2}
				+ g_0 |\psi_0|^2 \right)
		\right] \psi_0 \,.
	\label{GP}
    \end{alignat}
The linear quantum fluctuations $\hat \chi$ obey the Bogoliubov-de
Gennes equation \cite{Bogoliubov}
    \begin{alignat}{2}
	i b^2 \frac{\partial}{\partial t} \hat \chi \, = & \,
		\left[ - \frac{\bm \nabla_x^2}{2}
			+ f^2 \left(
				\frac{\bm x^2}{2}
				+ 2 g_0 |\psi_0|^2
			\right)
		\right] \hat \chi
		+ f^2 g_0 \psi_0^2 \hat\chi^\dagger \,,
	\label{BdG}
    \end{alignat}
and the residual terms comprise the equation of motion for $\hat \zeta$
    \begin{alignat}{2}
	i \frac{b^2}{f^2} \frac{\partial}{\partial t} \hat \zeta \, & = \,
		\left[
			- \frac{\bm \nabla_x^2}{2 f^2}
			+ \frac{\bm x^2}{2}
				+ 2 g_0 |\psi_0|^2
		\right] \hat \zeta
		+ g_0 \psi_0^2 \hat \zeta^\dagger
	\nonumber\\
	& \quad + g_0 \left( 2 \psi_0 \hat \chi^\dagger \hat \chi
			+ \psi_0^* \hat \chi^2
			+ \hat \chi^\dagger \hat \chi^2
		\right) + \ord( g_0 \hat \zeta) \,.
	\label{zeta}
    \end{alignat}
These higher orders $\hat \zeta$ must remain small in order for the
mean-field expansion \eqref{meanfield} to be valid, i.e., for the
linearized equation \eqref{BdG} to be applicable.
This means that the terms involving products of $\hat \chi$ must remain
small because they act as source terms for higher orders $\hat \zeta$.

From the Gross-Pitaevskii equation \eqref{GP}, I can infer 
when the evolution of the order parameter $\psi_0$ is solely described by
the scale factor $b(t)$:
apart from the trivial case $f^2 = \rm const$, this occurs only when
the spatial derivatives can be neglected with respect to the interaction
and trapping terms,
$\na_{\bm x}^2\psi_0 \ll f^2(\bm x^2 + 2 g_0 |\psi_0|^2)\psi_0$,
i.e., in the Thomas-Fermi approximation.
Density and phase of the order parameter
$\psi_0 = e^{i\phi_0}\sqrt{\varrho_0}$ then assume the form
    \begin{alignat}{2}
	\varrho_0^{\rm TF} \, & = \,
		\frac{\mu_0 - \bm x^2/2}{g_0} \Theta(\mu_0 - \bm x^2/2) \,,
	\nn
	\phi_0^{\rm TF} \, & = \,
		- \mu_0 \int\limits^t dt' \frac{f^2(t')}{b^2(t')} \,.
	\label{TF}
    \end{alignat}
where the Heaviside step function $\Theta(\mu_0 - \bm x^2/2)$ is $1$
for $\mu_0 > \bm x^2/2$ and $0$ elsewhere.
In this approximation, the motion of the classical background becomes
stationary and the scaled coordinates $\bm x$ are co-moving with the
condensate.

The Bogoliubov-de Gennes equation \eqref{BdG} can be tackled by introducing
self-adjoint operators
    \begin{alignat}{2}
	\hat \chi_+ \, & = \,
		e^{-i\phi_0} \hat\chi + e^{i\phi_0}\hat\chi^\dagger
	\,,\nonumber\\
	\hat \chi_- \, & = \,
		\frac{1}{2i}
		\left( 
			e^{-i\phi_0}\hat\chi - e^{i\phi_0}\hat\chi^\dagger
		\right) \,,
	\label{chipm}
    \end{alignat}
with $\phi_0 = \arg \psi_0$ being the phase of the order parameter.
These operators resemble (relative) density and phase fluctuations
$\delta\hat\varrho/\varrho_0 = \hat \chi_+ /\sqrt{\varrho_0}$ and
$\delta\hat\phi = \hat \chi_-/ \sqrt{\varrho_0}$ up to the prefactor
$1/\sqrt{\varrho_0}$.
Since this prefactor eventually becomes large near (and beyond) the surface
of the condensate, the smallness of $\delta\hat\varrho/\varrho_0$ and
$\delta\hat\phi$ cannot be ensured.
Therefore, I will stick to $\hat \chi_{\pm}$ in the following, but still
refer to them as density and phase fluctuations.
%
They obey
    \begin{alignat}{2}
	& 2 \left[ b^2 \frac{\partial}{\partial t}
		+\bm v_0\bm \nabla_x
		+ \frac{1}{2}(\bm \nabla_x \bm v_0)
	\right] \hat \chi_-
		\, = \, - \mathcal K_+ \hat \chi_+ \,,
	\nonumber\\
	& \frac{1}{2} \left[ b^2 \frac{\partial}{\partial t}
		+ \bm v_0 \bm \nabla_x
		+ \frac{1}{2} (\bm \nabla_x \bm v_0)
	\right] \hat \chi_+
		\, = \, \mathcal K_- \hat \chi_- \,,
	\label{eomchipm}
    \end{alignat}
where the velocity field $\vau_0 = \na_{\bm x} \phi_0$ results from
the residual background phase beyond the Thomas-Fermi approximation
\eqref{TF}.
However, $\vau_0$ is small for $g_0 > 0$ and if $f^2$ does not change too
swiftly, because an almost homogeneous phase will develop with only small
deviations near the boundary of the condensate.
Whereas attractive $g_0 < 0$ invalidate the Thomas-Fermi approximation
and generally $\vau_0 \neq 0$ even in the center of the trap.
The differential operators on the right-hand sides
    \begin{alignat}{2}
	\mathcal K_+ \, & = \,
		- \frac{\na_x^2}{2} + \frac{\vau_0^2}{2}
		+ f^2 \left( \frac{\bm x^2}{2} + 3 g_0 \varrho_0
			+ \frac{b^2}{f^2}\dot \phi_0 \right)
	\,,\nn
	\mathcal K_- \, & = \,
		- \frac{\na_x^2}{2} + \frac{\vau_0^2}{2}
		+ f^2 \left( \frac{\bm x^2}{2} + g_0 \varrho_0
			+ \frac{b^2}{f^2}\dot \phi_0 \right)
	\label{operatorsK}
    \end{alignat}
generally do not commute for inhomogeneous condensates
$\na_x \varrho_0 \neq 0$
    \begin{alignat}{2}
	\left[ \mathcal K_+, \mathcal K_- \right]
		\, & = \,
		f g_0 \left\{ (\na_x^2 \varrho_0)
			+ 2 (\na_x\varrho_0) \na_x
			\right\} \,.
	\label{commkpm}
    \end{alignat}
%


\section{Eigenmode expansion}
\label{SECeigenmode}


In order to define the initial state unambiguously, I will assume that the
condensate is at rest before $t = t_{\rm in}$.
Then $b = 1$, $\dot b = 0$, $f^2 = 1$, and $\vau_0 = 0$ such that the
left hand sides of Eqs.\ \eqref{eomchipm} reduce to partial time
derivatives and the initial eigenmode equations for $\hat \chi_\pm$
follow
%
%
    \begin{alignat}{2}
	\frac{\partial^2}{\partial t^2} \hat \chi_+
		\, & = \, - \mathcal K_- \mathcal K_+ \hat \chi_+
	\,,\nn
	\frac{\partial^2}{\partial t^2} \hat \chi_-
		\, & = \, - \mathcal K_+ \mathcal K_- \hat \chi_-
	\,.
    \end{alignat}
Because $\mathcal K_+ \mathcal K_- \neq \mathcal K_- \mathcal K_+$ for
inhomogeneous condensates, cf.\ Eq.\ \eqref{commkpm}, density and phase
fluctuations of each mode must have different space dependences.
This leads to the expansions \cite{inhom}
(I will adopt the sum convention throughout this Article for brevity;
any indices appearing only on one side of the equation are to be summed)
    \begin{alignat}{2}
	\hat \chi_+(\bm x, t) \, & = \, h_n^+(\bm x) \hat X_n^+(t) \,,
	\nn
	\hat \chi_-(\bm x, t) \, & = \, h_n^-(\bm x) \hat X_n^-(t) \,
	\label{eigenmode}
    \end{alignat}
of $\hat \chi_\pm$ into different eigenmode bases $\{h_n^+\}$ and
$\{h_n^-\}$, see Appendix \ref{APPeigenmodes} for more details on how
to obtain $h_n^\pm$.
Usually, these two bases are neither orthogonal nor normalized,
\mbox{$\int h_n^+ h_m^+ \neq \delta_{nm} \neq \int h_n^- h_m^-$},
but instead can be chosen to be dual to each other
    \begin{alignat}{2}
	\int d^Dx\, h_n^+(\bm x) h_m^-(\bm x) \, = \, \delta_{nm} \,.
	\label{dual}
    \end{alignat}
Note that this condition does not fix the norm of $h_n^\pm$ but still
permits the multiplication by an arbitrary factor,
$h_n^+ \to \Lambda_n h_n^+$ and $h_n^- \to (1/\Lambda_n) h_n^-$.
Observables must be unaffected by this ambiguity, see App.\
\ref{SECobservables}.

Insertion of the eigenmode expansion \eqref{eigenmode} into the linear
field equations \eqref{eomchipm} yields a set of coupled first-order
differential equations
    \begin{alignat}{2}
	b^2 \frac{\partial}{\partial t} \hat X_n^-
		\, & = \,
		- \frac12\mathcal A_{nm}(t) \hat X_m^+
		- \mathcal V_{nm}(t) \hat X_m^- \,,
	\nn
	b^2 \frac{\partial}{\partial t} \hat X_n^+
		\, & = \,
		2 \mathcal B_{nm}(t) \hat X_m^-
		+ \mathcal V_{mn}(t) \hat X_m^+ \,.
	\label{eigenmodeeom}
    \end{alignat}
with time-dependent coefficients.
The symmetric matrices
    \begin{alignat}{2}
	\mathcal A_{nm}(t) \, & = \,
		\int d^Dx\, h_n^+ \mathcal K_+ h_m^+ \,,
	\nn
	\mathcal B_{nm}(t) \, & = \,
		\int d^Dx\, h_n^- \mathcal K_- h_m^- \,,
	\label{coupl}
    \end{alignat}
are initially diagonal
$\mathcal A_{nm}(t_{\rm in}) = A_n(t_{\rm in}) \delta_{nm}$
and $\mathcal B_{nm}(t_{\rm in}) = B_n(t_{\rm in}) \delta_{nm}$.
At later times, they acquire off-diagonal elements because of the
different-time commutators
$\left[\mathcal K_+(t), \mathcal K_+(t')\right] \neq 0$ and
$\left[ \mathcal K_-(t), \mathcal K_-(t')\right] \neq 0$
when $\mathcal K_\pm(t) \neq \mathcal K_\pm(t')$.
The velocity coupling matrix
    \begin{alignat}{2}
	\mathcal V_{nm} \, & = \,
		\int d^Dx\, h_n^+\left[ \vau_0\na_x
			+ \frac{1}{2}(\na_x\vau_0)\right] h_m^-
	\label{velo}
    \end{alignat}
is not symmetric, but vanishes for homogeneous phases $\phi_0$ of the
order parameter, e.g., initially or in the Thomas-Fermi approximation
\eqref{TF}.
For repulsive $g_0 > 0$ and slow variations of interaction strength $g(t)$
and trap frequency $\omega(t)$, the order parameter phase is homogeneous
except for small ripples near the boundary of the condensate such that
the matrix $\mathcal V_{nm}$ is usually negligible.

From the evolution equations \eqref{eigenmodeeom} with the initially
diagonal coupling matrices \eqref{coupl}, the introduction of
bosonic operators $\hat a_n^\dagger$ and $\hat a_n$ creating or
annihilating an initial quasi-particle is straightforward
    \begin{alignat}{2}
	\hat X_n^-(t) \, & = \,
		F_n^m(t) \hat a_m + \bar F_n^m(t) \hat a_m^\dagger
	\,,\nn
	\hat X_n^+(t) \, & = \,
		G_n^m(t) \hat a_m + \bar G_n^m(t) \hat a_m^\dagger
	\,.
	\label{ineigenmodes}
    \end{alignat}
Here, a bar shall denote complex conjugation, e.g., $\bar F_n^m = (F_n^m)^*$.
The coefficients obey the initial values
    \begin{alignat}{2}
	F_n^m(t_{\rm in}) \, & = \,
		\sqrt{\frac{A_m}{2 \Omega_m}}
			\delta_{nm}
	\,,\,\,
	G_n^m(t_{\rm in}) \, & = \,
		i\sqrt{\frac{\Omega_m}{2 A_m}}
		\delta_{nm} \,.
	\label{eigenmodecoeff}
    \end{alignat}
where the phase has been appropriately chosen and the
frequencies $\Omega_m = \sqrt{A_mB_m}$.
The upper index of the coefficients $F_n^m$ and $G_n^m$ labels the mode,
while the lower index denotes the components of this particular mode when
expanded in a certain basis, e.g., the initial eigenfunctions $\{h_n^\pm\}$.

Since the coupling matrices \eqref{coupl} become non-diagonal
even for slow (adiabatically) variation of the trap frequency $\omega(t)$
or coupling strength $g(t)$, the initial bases $\{h_n^\pm\}$ cannot
represent the eigenmodes at later times.
Although $h_n^\pm$ might be employed in order to calculate the spatial
correlation functions, see Appendix \ref{SECobservables}, the use of
these functions might be misleading regarding the correlations between
different modes.
Furthermore, when probing the excitations using, e.g., the scheme
proposed in \cite{Ralf}, the proper particles defined at the time of
measurement will be detected and not the initial ones.

The particle definition in time-dependent background is a non-trivial
task, 
see, e.g., \cite{BirrellDavis}.
Nonetheless, it is always possible to expand $\hat \chi_\pm$
    \begin{alignat}{2}
	\hat \chi_\pm(\bm x,t) \, = \,
		h_{n;t_1}^\pm(\bm x) \hat X_{n;t_1}^\pm(t)
	\label{insteigenmode}
    \end{alignat}
into bases $\{h_{n;t_1}^\pm\}$, which are defined such that
the coupling matrices
\mbox{$\mathcal A_{nm;t_1} = \int d^Dx\,h_{n;t_1}^+ \mathcal K_+ h_{m;t_1}^+$}
and
\mbox{$\mathcal B_{nm;t_1} = \int d^Dx\,h_{n;t_1}^- \mathcal K_- h_{m;t_1}^-$},
cf.\ \eqref{coupl},
become diagonal at any particular instant $t_1$
    \begin{alignat}{2}
	\mathcal A_{nm;t_1}(t_1) \, & = \, A_{n;t_1}\delta_{nm}
	\,,\,\,\,
	\mathcal B_{nm;t_1}(t_1) \, & = \, B_{n;t_1}\delta_{nm} \,.
    \end{alignat}
Of course, the velocity term $\mathcal V_{nm;t_1}$ is then generally
non-diagonal and the evolution equations of the different modes will
not exactly decouple at this particular instant $t_1$.
But one should bear in mind that measurement occurs usually in an
adiabatic region, where the external parameters are only slowly-varying
functions of time.
Then, the background phase is approximately homogeneous and the velocity
$\vau_0 \approx 0$.
Hence, quasi-particle creators and annihilators might be introduced analogous
to Eq.\ \eqref{ineigenmodes}
    \begin{alignat}{2}
	\hat X_{n;t_1}^-(t_1) \, & = \,
		\sqrt{\frac{A_{n;t_1}}{2\Omega_{n;t_1}}}
		\left( \hat b_{n;t_1} + \,\hat b_{n;t_1}^\dagger \right)
	\,,\nn
	\hat X_{n;t_1}^+(t_1) \, & = \,
		i \sqrt{\frac{\Omega_{n;t_1}}{2A_{n;t_1}}}
		\left( \hat b_{n;t_1} - \,\hat b_{n;t_1}^\dagger \right) \,,
	\label{instmodes}
    \end{alignat}
where $\Omega_{n;t_1} = \sqrt{A_{n;t_1}B_{n;t_1}}$.



\section{Thomas-Fermi approximation and effective spacetime}
\label{SECTF}

In the previous sections, I made no approximations except for the
linearization \eqref{meanfield} and the assumption of an isotropic trap.
The formalism is, in principle, applicable for arbitrary variations of
trap frequency $\omega(t)$ and interactions $g(t)$.
To this end, it would be necessary to solve the Gross-Pitaevskii equation
\eqref{GP} and the linear evolution equations \eqref{eigenmodeeom}
simultaneously.
The numerical solution is complicated by the fact that the coupling matrices
\eqref{coupl} and \eqref{velo} need to be calculated at each time step.
Some of the numerical difficulties can be circumvent by adopting the
Thomas-Fermi profile \eqref{TF}, where density and phase of the background
become time-independent (in the coordinates $\bm x$) and thus require the
calculation of the coupling matrices only once.
Despite some shortcomings regarding the dynamics of the order parameter,
this approximation is usually applicable for repulsive interactions and
in the center of the trap, but becomes inaccurate towards the surface of
the condensate, where the quantum pressure $\propto (\na\sqrt{\varrho_0})^2$
is relevant.
%
%


\subsection{Coupled evolution equations}

Within the Thomas-Fermi approximation \eqref{TF}, the coordinate
transformation $\bm r \to \bm x$ associated with the scaling transformation
\eqref{scaling}
renders the background density time-independent
$\dot \varrho_0^{\rm TF} = 0$, while the phase becomes homogeneous
$\na_{x} \phi_0^{\pm TF} = 0$ and thus $\mathcal V_{nm}^{\rm TF} = 0$.
The integrals of the coupling matrices \eqref{coupl} simplify
considerably and the evolution equations can be cast into the form
    \begin{alignat}{2}
	- 2 b^2 \frac{\partial}{\partial t} \hat X_n^-
		\, & = \,
		A_n \hat X_n^+ + (f^2 - 1) \mathcal M_{nm} \hat X_m^+
	\,,\nn
	\frac12 b^2\frac{\partial}{\partial t} \hat X_n^+
		\, & = \,
		B_n \hat X_n^- + (f^2 - 1) \mathcal N_{nm} \hat X_m^- \,,
	\label{eigenmodeeomTF}
    \end{alignat}
i.e., $\mathcal A_{nm}$ and $\mathcal B_{nm}$ can be split into
time-independent diagonal parts,
cf.\ Eqs.\ \eqref{coupl},
    \begin{alignat}{2}
	A_n \delta_{nm} \, & = \, \int d^Dx \, h_n^+
		\left( -\frac{\na_{x}^2}{2}
			+ \frac{\bm x^2}{2} + 3 g_0\varrho_0^{\rm TF}
			- \mu_0
		\right) h_m^+
	\,,\nn
	B_n \delta_{nm} \, & = \, \int d^Dx\, h_n^-
		\left( -\frac{\na_{x}^2}{2}
			+\frac{\bm x^2}{2} + g_0\varrho_0^{\rm TF}
			- \mu_0
		\right) h_m^-
	\label{couplTFdiag}
    \end{alignat}
and constant, symmetric, non-diagonal coupling matrices
    \begin{alignat}{2}
	\mathcal M_{nm} \, & = \,
		\int d^Dx\, h_n^+ \left( \frac{\bm x^2}{2}
				+ 3 g_0\varrho_0^{\rm TF} - \mu_0
			\right) h_m^+
	\,,\nn
	\mathcal N_{nm} \, & = \,
		\int d^Dx\, h_n^- \left( \frac{\bm x^2}{2}
				+ g_0\varrho_0^{\rm TF} - \mu_0
			\right) h_m^-
	\label{couplTF}
    \end{alignat}
with time-dependent prefactors $f^2(t) - 1$.
The external variation of trap frequency $\omega(t)$ and coupling strength
$g(t)$ is solely encoded in the scale factor $b(t)$ and the scalar function
$f^2(t) = g(t) b^{2-D}$.
Note also that the coefficients \eqref{couplTF} and thus also the evolution
equations \eqref{eigenmodeeomTF} for the fluctuations are independent of
the initial coupling strength $g_0$.
The addend $g_0\varrho_0^{\rm TF}$ appearing in the parentheses of Eqs.\
\eqref{couplTF} and \eqref{couplTFdiag} can be expressed through the chemical
potential
$g_0 \varrho_0^{\rm TF} = (\mu_0 - \bm x^2 /2 ) \Theta(x_{\rm TF} - x)$,
where $x_{\rm TF} = \sqrt{2\mu_0}$, cf.\ the Thomas-Fermi equation
\eqref{TF}.


\subsection{Low energies and effective spacetime metric}

\label{SECspacetime}

The mode functions $h_n^\pm$ of excitations with low energies,
$\Omega_n\ll\mu_0$, are localized inside the condensate.
%
Hence, changing the bounds of the integrals \eqref{couplTFdiag} and
\eqref{couplTF} from infinity to the Thomas-Fermi radius $x_{\rm TF}$
will not alter these matrix elements significantly.
Bearing further in mind that
\mbox{$\bm x^2/2 + g_0\varrho_0^{\rm TF} - \mu_0 = 0$}
for $x < x_{\rm TF}$, it follows
$\mathcal N_{nm} = 0$ and
\mbox{$\mathcal M_{nm} = \int d^Dx\, h_n^+ 2 g_0\varrho_0 h_m^+$}.
Also
\mbox{$A_n\delta_{nm} = \int d^Dx\, h_n^+ 2g_0\varrho_0 h_m^+$}
because of the restriction to low energies,
\mbox{$\Omega_n = \sqrt{A_nB_n} \ll \mu_0 \approx g_0\varrho_0$}, and one
gets $\mathcal M_{nm} = A_n \delta_{nm}$.
Hence, the evolution equations of different eigenmodes approximately
decouple 
and I obtain second-order equations of motion for
phase \cite{densityandphasemodefunctions}
    \begin{alignat}{2}
	\left[\frac{\partial^2}{\partial\tau^2}
		- 2\frac{\partial \ln f}{\partial \tau}
		\frac{\partial}{\partial \tau}
		+ f^2 A_n B_n \right]
		\delta\hat \phi_n \, = \, 0 \,,
	\label{effKFGmode}
    \end{alignat}
and density fluctuations
    \begin{alignat}{2}
	\left[ \frac{\partial^2}{\partial\tau^2}
		+ f^2 A_n B_n \right]
		\delta\hat\varrho_n \, = \, 0 \,,
	\label{effKFGdensitymode}
    \end{alignat}
where I also introduced proper time $d\tau = dt/b^2$.
Equation \eqref{effKFGmode} is the evolution equation of a mode of a
minimally-coupled massless scalar field in a
Friedman-Lema\^\i tre-Robertson-Walker spacetime \cite{BirrellDavis,LL},
provided the scale factor
$a_{\rm FLRW}$ of the space-time is identified with $1/f$, cf.\ \cite{PeterF}
    \begin{alignat}{2}
	a_{\rm FLRW} \, = \, \frac{1}{f} \,.
	\label{scalefactor}
    \end{alignat}
[Note that the prefactor of the damping term is $2$ in Eq.\ \eqref{effKFGmode},
while it is $D$ in a $D+1$-dimensional Friedman-Lema\^\i tre-Robertson-Walker
spacetime.]
%



The analogy \eqref{effKFGmode} is not restricted to the evolution equations
in mode expansion but applies in the low-energy limit of the field equations
\eqref{eomchipm} as well:
within the Thomas-Fermi approximation,
$\bm x^2/2 + g_0\varrho_0 - \mu_0 = 0$ for $x < x_{\rm TF}$ and $\vau_0 = 0$,
and for low excitations,
$\na_{x}^2 \hat \chi_+ \ll 4 g_0\varrho_0 \hat \chi_+$,
the phase fluctuations obey a second-order field equation
\cite{PeterF}
    \begin{alignat}{2}
	\left[ \frac{\partial^2}{\partial \tau^2}
		- 2 \frac{\partial\ln f}{\partial \tau}
			\frac{\partial}{\partial\tau}
		- f^2 g_0\varrho_0 \na_{x}^2
	\right] \delta\hat\phi \, = \, 0 \,,
	\label{effKFG}
    \end{alignat}
which is similar to that of a minimally coupled scalar field
in a Friedman-Lema\^\i tre-Robertson-Walker spacetime.
%

Having established this kinematical analogy, cf.\ \eqref{effKFG}, some
of the concepts of general relativity can be applied to time-dependent
Bose-Einstein condensates.
%
Sonic analogs of horizons \cite{Wald,MTW,LL,Visser,HawkingEllis,RalfHor,MattHor}
are of particular interest for the study of non-equilibrium
effects, because they give a rough estimate whether and when adiabaticity
will be violated and the (quantum) fluctuations freeze and get amplified,
i.e., (quasi-)particle production occurs.
An effective particle horizon occurs, if a phonon emitted at a time $\tau_0$
can only travel a finite (co-moving) distance, i.e., if the integral
    \begin{alignat}{2}
	\Delta \, = \, \int\limits_{\tau_0}^\tau c_s(\tau') d\tau'
		\, = \, \sqrt{g_0\varrho_0}
			\int\limits_{\tau_0}^\tau f(\tau') d\tau'
	\label{horizon}
    \end{alignat}
converges to a finite value $\Delta_{\rm Horizon}$ for $\tau(t \to \infty)$.
Wavepackets emitted at time $\tau_0$ at the origin $\bm x = 0$ can reach
only points within the horizon, $x < \Delta_{\rm Horizon}$, in a finite
time.
All other points are concealed by the horizon.
[For simplicity, I assumed in Eq.\ \eqref{horizon} an homogeneous sound
velocity $c_s$.]


\subsection{Particle production in static traps}
\label{SECsweepstatic}

In order to point out the analogy of phase fluctuations to cosmic quantum 
fields, I formulated the evolution equations \eqref{effKFGmode} and
\eqref{effKFG} using proper
time $\tau$.
On the other hand, experiments are usually performed in the laboratory and
thus the variations of trap frequency $\omega$ and coupling strength $g$
are prescribed in laboratory time $t$.
Since $\tau$ is a complicated function of $t$, it is not quite obvious
whether or not the quantum fluctuations will experience
non-adiabatic evolution for a given modulation of $\omega(t)$ or $g(t)$.
In laboratory time $t$, Eq.\ \eqref{effKFGdensitymode} reads
    \begin{alignat}{2}
	\left[ \frac{\partial^2}{\partial t^2}
		+ 2 \frac{\dot b(t)}{b(t)} \frac{\partial}{\partial t}
		+ \Omega_n^2(t)
		\right]
		\delta\hat\varrho_n \, = \, 0 \,,
	\label{effKFGdensitymodet}
    \end{alignat}
which is the evolution equation of a damped harmonic oscillator with
time-dependent coefficients $2\dot b/b$ and
    \begin{alignat}{2}
	\Omega_n^2(t)
		\, = \, \left(\omega^2 + \frac{\ddot b}{b}\right) A_n B_n
		\, = \, \frac{g(t)}{b^{2+D}(t)} A_n B_n \,.
	\label{hydroinstfreq}
    \end{alignat}
Initially, when $\dot b = 0$, the field modes perform free oscillations.
Upon the gradual increase of the damping term $2 \dot b/b$ with respect
to the oscillation frequencies $\Omega_n$, the non-adiabatic evolution
of the quantum fluctuations slowly sets in, until they finally freeze and
get amplified when both terms, $2\dot b/b$ and $\Omega_n$ are of the
same order \cite{Review,Uwe2,Piyush,Silke,Barcelo,ich_njp,Uwe,PeterF}.

Let me discuss the two extremal ways a time-dependent scale factor
$b(t)$ can be achieved:
firstly, only the trap frequency might be varied, while the interaction
strength $g = 1$.
With instantaneous frequencies $\Omega_n \propto b^{-1-D/2}$, adiabaticity
can be violated for any finite change $\dot b/b \neq 0$ if $b$ becomes
sufficiently large.
The quantum fluctuations cannot adapt to the changing background any more,
they  freeze and get amplified.
And, secondly, for static traps, $\omega = 1$, where $g(t)$ is time-dependent.
%
Then, the situations is not so clear because
$\Omega_n^2 \propto g/b^{2+D} = 1 + \ddot b /b$, cf.\ Eq.\ \eqref{fvont}.
Hence, only the rapid acceleration of the scale factor
$|\ddot b/b| \gtrsim \ord(1)$ will lead to notable changes of the
excitation frequencies.
On the other hand, adiabaticity could be violated by increasing the
magnitude of the damping term, $2\dot b/b$.
Then, however, a continuous acceleration of $b$ is required, because
otherwise, if $\dot b/b$ was constant, the system would equilibriate.
%


As an example for the absence of particle production inside a static
trap, $\omega = 1$, I will consider an exponential sweep of the coupling
coefficient
    \begin{alignat}{2}
	g(t) \, = \, \exp\{ \gamma t \} \,
	\label{expsweepg}
    \end{alignat}
with $\gamma > 0$.
In this case, Eq.\ \eqref{fvont} for the scale factor permits an analytic
solution
    \begin{alignat}{2}
	b(t) \, = \, \left[ \frac{(2+D)^2}{\gamma^2 + (2+D)^2}
		\right]^{\frac1{2+D}} \exp\left\{\frac{\gamma t}{2+D}\right\} \,.
	\label{bexpsweepg}
    \end{alignat}
The coefficients of Eq.\ \eqref{effKFGdensitymodet} become time-independent
    \begin{alignat}{2}
	2 \frac{\dot b}{b} \, = \, \frac{2 \gamma}{2 + D}
	\,,\quad
	\Omega_n^2
	\, = \, \left( 1 + \frac{\gamma^2}{(2+D)^2}\right) A_n B_n \,,
    \end{alignat}
and the density eigenmodes are just damped harmonic oscillators with
solutions
    \begin{alignat}{2}
	\delta\hat\varrho_n \, = \,
		e^{-\gamma t/(2 + D)} \delta\hat\varrho_n' \,,
    \end{alignat}
where $\delta\hat\varrho_n'$ is some residual oscillating function with
frequency $\Omega_n$.
Hence, the density-density fluctuations diminish
$\langle (\hat X_{n;t}^+)^2\rangle \propto e^{- 2 \gamma t/(2+D)}$, and,
consequently, the phase-phase fluctuations increase.
But this is just the adiabatic evolution, because
$A_{n;t} \propto e^{2\gamma t/(2+D)}$ and thus
$\langle \hat X_{n;t}^+ \rangle_{\rm ad} = \Omega_{n;t}/(2A_{n;t})
\propto e^{-2\gamma t/(2+D)}$, cf.\ App.\ \ref{SECobservables}.
This means that no quasi-particle production occurs for the dynamics
\eqref{expsweepg} in the hydrodynamic regime, i.e., for low-energy
excitations with order parameter treated in the Thomas-Fermi
approximation.
These findings can also be inferred from the (absence of an) effective
particle horizon \eqref{horizon}.
For the particular shape \eqref{bexpsweepg} of the scale factor $b$ follows
    \begin{alignat}{2}
	\Delta \, \propto \, \int\limits_0^\infty
		f(t) \frac{dt}{b^2(t)}
		\, = \, \int\limits_0^\infty
		dt \to \infty \,,
	\label{horexpsweep}
    \end{alignat}
because $f(t) \propto \exp\{ 2 \gamma t / ( 2 + D) \} \propto b^2(t)$.

Note that the presented solution
assumes $g = e^{\gamma t}$ at all times, especially also for
$t < t_{\rm in}$.
Hence $g(t') \neq 1$ at some time $t'$ when $b(t') = 1$,
cf.\ Eqs.\ \eqref{expsweepg} and \eqref{bexpsweepg}.
On the other hand, $b(t_{\rm in}) = 1$ and $g(t_{\rm in}) = 1$
for a condensate at rest, see Eq.\ \eqref{fvont}.
Since both solutions (static initial state and exponential sweep) cannot
be matched at $t_{\rm in}$ such that $g$ and $b$ are both continuous, the
switching on of the exponential sweep would excite breathing oscillations.
These oscillations, however, generally affect particle production,
e.g., through parametric resonance.


\section{Quasi-one-dimensional condensate}
\label{SEC1D}

The simplest application of the presented formalism consists
in a quasi-one-dimensional condensate.
In highly anisotropic traps, where the perpendicular trap frequency
$\omega_\perp$ is much larger than the chemical potential, the motion
in the perpendicular directions is restricted to the ground state and
might be integrated out.
An effectively one-dimensional field equation \eqref{fieldequation} follows,
where the interaction strength
    \begin{alignat}{2}
	g_{\rm 1D} \, = \, \frac{g_{\rm 3D}}{2 \pi a_\perp^2}
		\, = \, 2 a_s \omega_\perp
    \end{alignat}
can be varied through Feshbach resonance or by changing $\omega_\perp$.
However, one should bear in mind that the transversal extent of the
condensate $a_\perp = 1/\sqrt{m\omega_\perp}$ has to be much larger than
the $s$-wave scattering length $a_s$ such that the interaction of different
atoms can still be described through three-dimensional scattering theory.
For simplicity, I will adopt in this section the Thomas-Fermi approximation
\eqref{TF} for the order parameter but will permit arbitrary energies
for the excitations.


\subsection{Spectrum}

    \begin{figure}[!hbt]
	\includegraphics[width=0.4\textwidth]
		{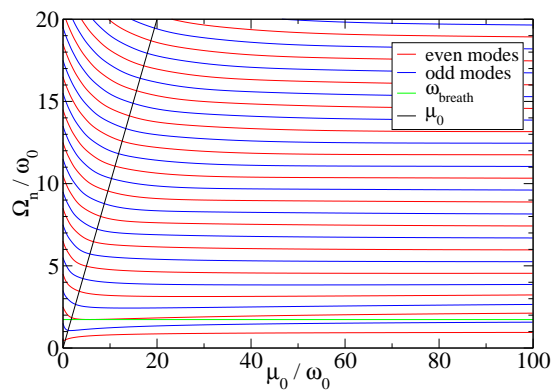}
	\caption{Frequencies of odd (blue) and even (red) excitations
		versus the chemical potential $\mu_0$ calculated using
		100 harmonic oscillator basis functions.
		The frequencies equal the chemical potential at the
		intersection with the black line.
		The lowest odd mode converges for $\mu_0 \to \infty$ to
		the classical Thomas-Fermi breathing frequency
		$\Omega_{\rm breath}^{\rm TF} = \sqrt{3}\omega_0$ (green),
		which follows from Eq.\ \eqref{fvont}.
		}
	\label{spectrum1d}
    \end{figure}

In Figure \ref{spectrum1d}, the excitation frequencies
$\Omega_n = \sqrt{A_n B_n}$ of the lowest modes are plotted
versus the chemical potential $\mu_0$.
For $\mu_0 = 0$, one has the equidistant spectrum of the harmonic
oscillator $\Omega_n = (n + 1/2) \omega_0$.
Whereas for high $\mu_0 \gg \Omega_n$, the frequencies become almost
independent of the chemical potential.
In particular, the frequency $\Omega_1$ of the lowest excitation with
odd parity tends for $\mu_0 \to \infty$ towards the Thomas-Fermi breathing
frequency $\Omega_{\rm breath}^{\rm TF} = \sqrt{3} \omega_0$ obtained from
Eq.\ \eqref{fvont}.

The discrepancy between these two frequencies $\Omega_1$ and
$\Omega_{\rm breath}^{\rm TF}$ for finite values of the chemical potential
$\mu_0$ hints at shortcomings of the Thomas-Fermi approximation.
In particular, Eq.\ \eqref{fvont} does not describe the breathing motion
of the background properly and care must be taken when employing
Eqs.\ \eqref{TF} for the order parameter.
Nonetheless, Eq.\ \eqref{fvont} still predicts the correct order of
magnitude of the characteristic response time of the background,
$1/\omega_{\rm breath} = \ord(1/\omega_{\rm breath}^{\rm TF})$,
to variations of trap frequency or interaction strength.

Hence, it is still possible to discuss several cases, where the difference
of $\omega_{\rm breath}$ and $\omega_{\rm breath}^{\rm TF}$ is either
small or does not matter:
firstly, if the shape of the condensate varies only slowly, i.e., if
$\dot b \ll \omega_{\rm breath}$ and the condensate can adapt to
changes of $\omega$ and/or $g$ immediately.
Secondly, if no breathing oscillations are excited, e.g., because the
condensate expands or contracts, see also Eqs.\ \eqref{expsweepg}
and \eqref{bexpsweepg}.
And, thirdly, for very large chemical potentials, $\mu_0 \to \infty$, the
Thomas-Fermi approximation becomes exact.
The quantum fluctuations are in the hydrodynamic regime and their
effective evolution equations \eqref{effKFGmode} decouple.
In order to address cases where breathing of the background occurs,
it would be necessary to abandon the Thomas-Fermi approximation \eqref{TF}
and to solve the Gross-Pitaevskii equation \eqref{GP}, which could, e.g.,
be done by expanding the order parameter $\psi_0$ into oscillator functions
\cite{GP-modeexp}.


\subsection{Exponential sweep in stationary condensate}

\label{sweepstationary}

A stationary condensate, i.e., $b = 1$, can be accomplished through
simultaneous variation of trap frequency $\omega$ and interaction strength
$g$, cf.\ Eq.\ \eqref{fvont}
    \begin{alignat}{2}
	f^2(t) \, = \, \frac{\omega^2(t)}{\omega_0^2}
		\, = \, \frac{g(t)}{g_0} \,,
	\label{simultan}
    \end{alignat}
though, one should be aware that this only holds within the
Thomas-Fermi approximation:
the instantaneous chemical potential must at all times be much larger than
the trap frequency
    \begin{alignat}{2}
	\mu_{\rm inst}(t) \, = \, \mu_0 f^2(t)
		\gg \omega(t) \, = \, \omega_0 f(t) \,.
	\label{instTF}
    \end{alignat}
If both were of the same order, the kinetic term, $-\na_{\bm x}^2\psi_0/2$,
in the Gross-Pitaevskii equation \eqref{GP} becomes important;
no stationary background could be realized even for simultaneous
variation of $\omega$ and $g$.


\subsubsection{Analytical effective spacetime solution}

For an exponential sweep
    \begin{alignat}{2}
	f^2(t) \, = \, e^{- 2 \gamma t} \,,\qquad
	\gamma \, > \, 0 \,,
	\label{expsweep}
    \end{alignat}
the effective second-order equation \eqref{effKFG} for the phase
fluctuations in the hydrodynamic limit
    \begin{alignat}{2}
	\left[
		\frac{\partial^2}{\partial t^2}
		+ 4 \gamma \frac{\partial}{\partial t}
		- f^2 g_0 \varrho_0 \na_x^2
	\right] \delta\hat\phi \, = \, 0 \,.
	\label{effKFGexp}
    \end{alignat}
is that of a massless scalar field in a de~Sitter
spacetime with exponentially growing scale factor
$a_{\rm FLRW} = 1/f = e^{\gamma t}$, cf.\ Eq.\ \eqref{scalefactor}
-- which is believed to describe the universe during the
epoch of cosmic inflation \cite{LL,Wald}.
For the time-dependence \eqref{expsweep}, the integral \eqref{horizon}
is finite and an effective sonic horizon occurs;
the quantum fluctuations freeze and get amplified.

Instead of solving Eq.\ \eqref{effKFGexp} for 
$\delta \hat \phi$,
I will consider the evolution equation of the density fluctuations
$\hat X_n^+$
    \begin{alignat}{2}
	\left [\frac{\partial^2}{\partial t^2}
		+ \gamma^2 e^{-2 \gamma (t - t_n)}
	\right] \hat X_n^+ \, = \, 0 \,,
	\label{densitysweepeom}
    \end{alignat}
where $A_nB_n = \gamma^2 e^{2 \gamma t_n}$.
Obviously, all modes undergo the same evolution just at different times.
Eq.\ \eqref{densitysweepeom} can be solved analytically in terms of Bessel
functions \cite{Abramowitz}
    \begin{alignat}{2}
	\hat X_n^+ \, = \,
		\sqrt{\frac{\pi B_n}{2\gamma}}
			\left\{ \hat a_n H_0^{(1)}(e^z)
				+ \hat a_n^\dagger H_0^{(2)}(e^z)
			\right\} \,.
	\label{densitysweep}
    \end{alignat}
where $z = - \gamma(t - t_n)$.
The Hankel functions $H_0^{(1/2)}$ have the proper asymptotics for
early times $t \to - \infty$ such that the operators $\hat a_n$ annihilate
the initial vacuum state.
The phase fluctuations
\mbox{$\hat X_n^- = (1/2B_n) \partial \hat X_n^+/\partial t$} read
    \begin{alignat}{2}
	\hat X_n^- \, = \,
		\sqrt{\frac{\pi\gamma}{8B_n}}
		\left\{
			\hat a_n e^z H_1^{(1)}(e^z)
			+ \hat a_n^\dagger e^z H_1^{(2)}(e^z)
		\right\} \,.
	\label{phasesweep}
    \end{alignat}
From these expressions \eqref{densitysweep} and \eqref{phasesweep},
I can infer the correlations of each mode.
At late times follows
    \begin{alignat}{2}
	\left\langle (\hat X_n^+)^2 \right\rangle(t \to \infty)
		\, & = \,
		\frac{2\gamma B_n}{\pi}(t - t_n)^2
	\,,\nn
	\left\langle (\hat X_n^-)^2 \right\rangle(t\to\infty)
		\, & = \,
		\frac{\gamma}{2\pi B_n} \,.
	\label{corrsweep}
    \end{alignat}
Comparison with the adiabatic values
$\langle(\hat X_n^\pm)^2\rangle_{\rm ad}$, cf.\
Eq.\ \eqref{instpartnum},
yields the quasi-particle number
at late times
    \begin{alignat}{2}
	N_n(t \to \infty ) \, = \,
		\frac{1}{\pi} e^{\gamma(t - t_n)} - 1 \,.
	\label{nsweep}
    \end{alignat}
The occupation number of all modes grows exponentially though at different
times $t - t_n$, where the shift $t_n$ is determined by the excitation
frequencies $\sqrt{A_nB_n} = \gamma e^{\gamma t_n}$.
However, one should bear in mind that Eq.\ \eqref{densitysweepeom} is only
valid for a limited time before leaving the hydrodynamic regime.
%


\subsubsection{Numerical results}

In order to go beyond the effective space-time description and thus the
analytical findings \eqref{densitysweep}-\eqref{nsweep},
I will now consider the full evolution equations
\eqref{eigenmodeeomTF}.
The sweep rate $\gamma = 0.1$ shall be chosen such that all modes
evolve adiabatically at first, i.e., $\gamma \ll \Omega_n(t_{\rm in})$
for all $n$.
When subsequently reducing trap potential and coupling strength, the
excitation frequencies $\Omega_n(t)$ decrease and non-adiabatic evolution
sets in at different times for each mode.
The fluctuations freeze and get amplified.
%
%

\begin{figure}[!hbt]
	\includegraphics[width=0.4\textwidth]{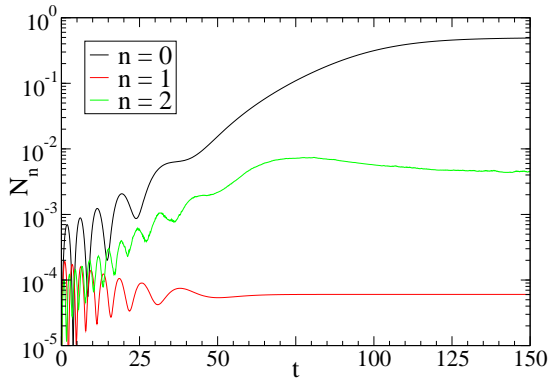}
	\caption{Instantaneous particle number in the lowest three
		quasi-particle modes for an exponential sweep
		\eqref{expsweep}.
		The initial chemical potential is $\mu_0 = 50$
		and the sweep rate $\gamma = 0.1$.
		At $t \approx 80$, the chemical potential equals the
		trap frequency.
		For the calculation of the eigenmodes $h_n^\pm$ and the
		coupling matrices \eqref{coupl}, I used the harmonic
		oscillator functions $h_\alpha(x)$ up to $\alpha = 79$,
		cf.\ Appendix \ref{SECeigenmode}.
		The lowest 20 modes $\hat X_n^\pm$ were then propagated.
		The numerical accuracy was set to $10^{-6}$.
		}
	\label{sweep50}
    \end{figure}

Figure \ref{sweep50} shows the instantaneous particle numbers of the lowest
three modes for initial chemical potential $\mu_0 = 50$ and
$\gamma = 0.1$.
The lowest excitation, $n = 0$, which becomes non-adiabatic first,
acquires the largest particle number.
The next two modes, $n = 1, 2$, experience less squeezing, 
though, remarkably, $N_2 > N_1$ -- an unexpected result, which can be
explained by the coupling of different modes:
the second even mode, $n = 2$, gets populated from the principle excitation,
$n = 0$, whereas the coupling matrix elements between $n = 0$ and $n = 1$
are zero because of different parity.
%
%
%




\section{Summary}
\label{SECsum}

The main objective of this Article was the investigation of quantum
fluctuations in time-dependent harmonically-trapped Bose-Einstein
condensates with repulsive interactions.
To this end, the linear fluctuations were expanded into their initial
eigenmodes and the field equations were diagonalized.
This diagonal form, however, persists only as long as the condensate is
at rest;
as soon as trap frequency or interaction strength are varied, the coupling
of different modes sets in.
(Only part of which can be accounted for by transforming to the instantaneous
eigenmodes, though the definition of instantaneous eigenmodes is a
non-trivial task.)

Two regimes were identified:
firstly, for energies much smaller than the chemical potential, the
coupling of different modes is negligible and an effective space-time
metric might be introduced for the phase fluctuations.
This, however, necessitates a redefinition of the time coordinate
such that the required change of trap frequency and/or interaction
strength for a certain dynamics of this effective space-time is not
obvious.
It turned out that the sole variation of the interaction coefficient
$g(t)$ in a smooth monotonic way is hardly sufficient to render the
evolution of the quantum fluctuations non-adiabatic, since the
expansion/contraction of the background might compensate for changes
of $g$ such that the sound velocity (in comoving coordinates) remains
constant.
Breathing oscillations of the background, excited, e.g., by the
sudden change of the interaction coefficient $g$, might still yield
a notable amount of quasi-particles.
And secondly, if the excitation energy is of the same order as the
chemical potential, different quasi-particle modes couple.

The amplification and freezing of the fluctuations and also the coupling
of different modes was illustrated in an example, where the trap frequency
and interaction strength were exponentially ramped down such that the
shape of the condensate remains constant.
For the considered parameters, a quasi-particle number of 0.5 was
obtained in the lowest mode, though higher occupation numbers could
be achieved by faster sweep rates or starting with a higher
chemical potential.
The inversion of the occupation number in the next two modes could
be attributed to the inter-mode coupling:
although the third excitation, $n = 2$, experiences a much briefer
period of non-adiabatic evolution than the second mode, $n = 1$,
only the former couples to the lowest mode, $n = 0$, and gets populated
from it.


\acknowledgments

I would like to thank Craig M.\ Savage and Uwe R.\ Fischer for helpful
discussions.
This research was supported by the Australian Research Council.

\appendix

\section{Eigenmodes}
\label{APPeigenmodes}

The aim of this Appendix is the derivation of the initial eigenfunctions
$h_n^\pm$ of density and phase fluctuations.
To this end, let me expand the field operator into any orthonormal basis
$\{h_\alpha(\bm x)\}$ of the underlying Hilbert space $L^2(\mathbb R^D)$
with some operator-valued coefficients $\hat \chi_\alpha^\pm(t)$
    \begin{alignat}{2}
	\hat \chi_\pm(\bm x,t) \, = \,
		h_\alpha(\bm x) \hat \chi_\alpha^\pm(t) \,.
	\label{basisexp}
    \end{alignat}
Lower-case Greek indices ($\alpha$, $\beta$,...) shall denote the
components in this arbitrary basis, while lower-case Latin indices
($m$, $n$,...) label the initial eigenmodes.

If the condensate is initially at rest, the scale factor \mbox{$b = 1$} and
the phase of the order parameter is homogenous $\na_{\bm x} \phi_0 = 0$
such that $\vau_0 = 0$ and thus $\mathcal V_{nm} = 0$, cf.\
\eqref{velo}.
The evolution equations \eqref{eomchipm} simplify considerably and can
be expanded into the basis $\{h_\alpha\}$.
It follows, cf.\ \eqref{eigenmodeeom}
    \begin{alignat}{2}
	- 2 \frac{\partial}{\partial t} \hat \chi_\alpha^-
		\, & = \, \mathcal A_{\alpha\beta} \hat \chi_\beta^+
		\, & = & \, \int d^Dx\,h_\alpha\mathcal K_+h_\beta
			\hat \chi_\beta^+
	\,,\nn
	\frac{1}{2} \frac{\partial}{\partial t} \hat\chi_\alpha^+
		\, & = \, \mathcal B_{\alpha\beta}\hat\chi_\beta^-
		\, & = & \, \int d^Dx\,h_\alpha\mathcal K_-h_\beta
			\hat \chi_\beta^-
	\,,
	\label{initialeombasis}
    \end{alignat}
where $\mathcal K_\pm$ are defined in Eq.\ \eqref{operatorsK}.
The matrices $\mathcal A_{\alpha\beta}$ and $\mathcal B_{\alpha\beta}$
are real and symmetric but do not commute due to
$\left[\mathcal K_+, \mathcal K_-\right] \neq 0$, cf.\ \eqref{commkpm}.
From Eqs.\ \eqref{initialeombasis}, I obtain second-order evolution
equations for phase and density fluctuations
    \begin{alignat}{2}
	\frac{\partial^2}{\partial t^2} \hat \chi_\alpha^-
		\, & = \,
			- \mathcal A_{\alpha\beta} \mathcal B_{\beta\gamma}
				\hat \chi_\gamma^-
	\,,\nn
	\frac{\partial^2}{\partial t^2} \hat \chi_\alpha^+
		\, & = \,
			- \mathcal B_{\alpha\beta} \mathcal A_{\beta\gamma}
				\hat \chi_\gamma^+ \,,
	\label{inchipm}
    \end{alignat}
which can be diagonalized by a transformation to the eigenvectors.

Because the matrix 
$\mathcal A_{\alpha\beta}\mathcal B_{\beta\gamma} \neq
\mathcal B_{\alpha\beta}\mathcal A_{\beta\gamma}$
is not symmetric,
phase and density fluctuations obey different eigenvalue equations
    \begin{alignat}{2}
	&\mathcal A_{\alpha\beta}\mathcal B_{\beta\gamma}&
		v_\gamma^n \, & = \, \lambda_n v_\alpha^n
	\,,\nn
	&\mathcal B_{\alpha\beta}\mathcal A_{\beta\gamma}&
		\tilde v_\gamma^n
	\, & = \, \tilde \lambda_n \tilde v_\alpha^n
	\,,
	\label{eigenab}
    \end{alignat}
where $n$ denotes the vectors and $\alpha$ merely counts the components
in the particular basis representation \eqref{basisexp}.
For brevity, I will only discuss the eigenvectors of $\mathcal A\mathcal B$
in the following, though the same applies for those of $\mathcal B\mathcal A$
as well.
The eigenvectors are generally not orthogonal 
    \begin{alignat}{2}
	\sum_\alpha v_\alpha^n v_\alpha^m
		\, & \neq \, \delta_{nm}
	\,,
    \end{alignat}
but $\{v_\alpha^n\}$ usually still forms a basis.
(Note, however, that the eigenvectors of a non-symmetric matrix not always
span the entire vector space.
But if the $v^n$ were no basis, the evolution equation \eqref{inchipm}
could not be diagonalized.
This would mean that there existed some fluctuations,
which have constant losses to some eigenmodes -- rather unphysical
in view of the stationary initial state considered here.
Therefore, I will not discuss this case any further.)
With the $v^n = h_\alpha v_\alpha^n$ forming a basis of $L^2(\mathbb R^D)$,
there must exist a dual basis $\{v^{{\rm d},n}\}$ in the space of
linear functionals (the dual) on $L^2(\mathbb R^D)$, which obeys
    \begin{alignat}{2}
	\int d^Dx\,v^{{\rm d},n} v^m \, = \,
	\sum_\alpha v_\alpha^{{\rm d},n} v_\alpha^m
		\, & = \, \delta_{nm} \,.
	\label{dualv}
    \end{alignat}
[Roughly speaking, the elements of $L^2(\mathbb R^D)$ become column vectors
in the basis expansion \eqref{basisexp}, whereas row vectors correspond to
the functionals on $L^2(\mathbb R^D)$, i.e., the elements of the dual.
Since the eigenvectors $v^n$ form a basis, the matrix with the $v^n$ as
columns must be invertible.
The rows of the (left) inverse matrix then comprise the elements of the
dual basis $\{v^{{\rm d},n}\}$ in the particular basis expansion
\eqref{basisexp}.]

After the multiplication of the first of Eqs.\ \eqref{eigenab}
with $v_\alpha^{{\rm d},m}$ from the left and summation over $\alpha$
follows
    \begin{alignat}{2}
	v_\alpha^{{\rm d},m}
		\mathcal A_{\alpha\beta}\mathcal B_{\beta\gamma}
		v_\gamma^n \, = \,
		v_\alpha^{{\rm d},m} v_\alpha^n \lambda_n
		\, = \, \delta_{nm} \lambda_m \,,
	\label{dualdiag}
    \end{alignat}
which, because $\{v^n\}$ is a basis,
implies that the $v^{{\rm d},n}$ are the left eigenvectors of
$\mathcal A_{\alpha\beta}\mathcal B_{\beta\gamma}$ with the same
eigenvalues $\lambda_n$
    \begin{alignat}{2}
	v_\alpha^{{\rm d},n} \mathcal A_{\alpha\beta}
		\mathcal B_{\beta\gamma}
		\, & = \, v_\gamma^{{\rm d},n} \lambda_n \,.
    \end{alignat}
Transposition yields
    \begin{alignat}{2}
	\mathcal B_{\gamma\beta}\mathcal A_{\beta\alpha}
		v_\alpha^{{\rm d},n} \, & = \,
		\lambda_n v_\gamma^{{\rm d},n} \,.
    \end{alignat}
i.e., the eigenvalue equation of $\mathcal B\mathcal A$,
cf.\ \eqref{eigenab}.
Hence, $\tilde v^n \propto v^{{\rm d},n}$ and
$\mathcal A \mathcal B$ and $\mathcal B \mathcal A$ (i.e., density
and phase fluctuations) must have the
same spectrum $\{\lambda_n\} = \{\tilde \lambda_n\}$.
For simplicity, $\tilde v^n = v^{{\rm d},n}$, which can be achieved
by renormalization of $\tilde v^n$, in the following.

Note that the spectrum of a real non-symmetric matrix might contain
pairs of complex conjugate eigenvalues $\lambda_n$, $\lambda_n^*$,
which can be seen when taking the complex conjugate of \eqref{eigenab}.
Complex eigenvalues $\Im \lambda_n \neq 0$ are associated with
exponentially growing solutions, i.e., unstable modes.
Since I am interested in the quantization of the stationary initial state,
I will not discuss this case but instead assume $\lambda_n \in \mathbb R$
$\forall n$.
(As can be easily verified, the components of the eigenvectors
$v_\alpha^n$ and $\tilde v_\alpha^n$ must be real-valued as well.)
Nonetheless, complex eigenvalues might still occur in dynamical
situations, e.g., during the signature change event proposed in
\cite{Silke} or during (quantum) phase transitions \cite{Sachdev}
(see also Ref.\ \cite{spinor} for an illustrative example).

The initial evolution equations \eqref{inchipm} can be diagonalized by
multiplication with the left eigenvectors
    \begin{alignat}{2}
	\frac{\partial^2}{\partial t^2} \tilde v_\alpha^n \hat \chi_\alpha^-
		\, & = \,
		- \tilde v_\alpha^n \mathcal A_{\alpha\beta}
			\mathcal B_{\beta\gamma}\hat\chi_\gamma^-
		\, & = & \, - \lambda_n \tilde v_\gamma^n\hat \chi_\gamma^-
	\,,\nn
	\frac{\partial^2}{\partial t^2} v_\alpha^n \hat \chi_\alpha^+
		\, & = \,
		- v_\alpha^n \mathcal B_{\alpha\beta}
			\mathcal A_{\beta\gamma}\hat\chi_\gamma^+
		\, & = & \, - \lambda_n v_\gamma^n\hat \chi_\gamma^+ \,.
    \end{alignat}
which leads to the definition of the density and phase fluctuation eigenmodes
$\hat X_n^\pm$ through
    \begin{alignat}{2}
	\hat X_n^- \, & = \, \tilde v_{\alpha}^n \hat \chi_\alpha^-
	\,,\quad
	& \hat X_n^+ \, & = \, v_{\alpha}^n \hat \chi_\alpha^+ \,,
	\label{oscillmode2}
    \end{alignat}
where the spatial mode functions
    \begin{alignat}{2}
	h_n^- \, = \, v_{\alpha}^n h_\alpha
	\,,\quad
	h_n^+ \, = \, \tilde v_{\alpha}^n h_\alpha \,.
    \end{alignat}
follow from comparison of Eqs.\ \eqref{basisexp} with \eqref{eigenmode}
and the duality \eqref{dualv} of the $v_\alpha^n$ and $\tilde v_\alpha^n$
implies Eq.\ \eqref{dual} for the $h_n^\pm$.

In view of \eqref{oscillmode2}, the initial evolution equations
\eqref{initialeombasis} can be transformed to the new basis, cf.\
Eq.\ \eqref{eigenmodeeom}
    \begin{alignat}{2}
	- 2\frac{\partial}{\partial t} \hat X_n^-
		\, & = \, A_n \hat X_n^+
	\,,\quad
	\frac{1}{2}\frac{\partial}{\partial t} \hat X_n^+
		\, & = \, B_n \hat X_n^-
    \end{alignat}
where the transformed matrices
    \begin{alignat}{2}
	\mathcal A_{nm} \, & = \,
		\tilde v_\alpha^n \mathcal A_{\alpha\beta} \tilde v_\beta^m
		\, && = \, A_n \delta_{nm} \,,
	\nn
	\mathcal B_{nm} \, & = \,
		v_\alpha^n \mathcal B_{\alpha\beta} v_\beta^m
		\, && = \, B_n \delta_{nm} \,.
	\label{diagcouplmatrix}
    \end{alignat}
have become diagonal.
Note that the transformation \eqref{oscillmode2} is not orthogonal
because the matrices comprising of the eigenvectors $v_\alpha^n$
and $\tilde v_\alpha^n$ are not orthogonal.
Hence, the commutators are {\em not} preserved, in particular
    \begin{alignat}{2}
	\mathcal A_{\alpha\beta}\mathcal B_{\beta\gamma}
		\, & \neq \,
		\mathcal B_{\alpha\beta}\mathcal A_{\beta\gamma}
	\,,\nn
	\mathcal A_{nm} \mathcal B_{ml}
		\, & = \, \mathcal B_{nm}\mathcal A_{ml}
    \end{alignat}
where the latter can be inferred from Eq.\ \eqref{dualdiag}
    \begin{alignat}{2}
	\lambda_n \delta_{nm} \, & = \,
		\tilde v_\alpha^n \mathcal A_{\alpha\beta}
		\tilde v_\beta^k \,
		v_\gamma^k
			\mathcal B_{\gamma\delta} v_\delta^m
		\, & = & \,
		\mathcal A_{nk}\mathcal B_{km} \, = \,
	\nn
	(\lambda_n\delta_{nm})^T
		\, & = \,
		v_\alpha^n \mathcal B_{\alpha\beta} v_\beta^k\,
		\tilde v_\gamma^k \mathcal A_{\gamma\delta}
			\tilde v_\delta^m
		\, & = & \, \mathcal B_{nk}\mathcal A_{km}
		\,.
    \end{alignat}
Since $\mathcal A_{nm}$ and $\mathcal B_{nm}$ also commute with their
product $\mathcal A_{nk}\mathcal B_{km} = \lambda_n \delta_{nm}$, which
is diagonal, they must be diagonal, too.


\section{Observables}
\label{SECobservables}

The duality condition \eqref{dual} does not fix the norm of the eigenvectors
but still permits the multiplication with an arbitrary factor $\Lambda_n$
    \begin{alignat}{2}
	h_n^+ &\to \Lambda_n h_n^+
	\,,\qquad
	&h_n^- &\to \frac{1}{\Lambda_n} h_n^- \,.
    \end{alignat}
This renormalization then leads to a stretching/shrinking of the
operator-valued coefficients, cf.\ \eqref{eigenmode}
    \begin{alignat}{2}
	\hat X_n^+ & \to \frac{1}{\Lambda_n} \hat X_n^+
	\,,\qquad
	&\hat X_n^- & \to \Lambda_n \hat X_n^- \,.
    \end{alignat}
Since this is merely a 
basis transformation,
the time-evolution of the quantum fluctuations $\hat X_n^\pm$ must be
unaffected.
To see this, recall the definitions \eqref{coupl} and \eqref{velo} of the
coupling matrices:
they are the matrix elements of the operators $\mathcal K_\pm$ and of
$\vau_0 \na_{\bm x} + (\na_{\bm x}\vau_0)/2$ with respect to the basis
functions $h_n^\pm$ and thus acquire additional factors as well.
As expected, all of these factors cancel such that the time-evolution
remains unchanged.
In particular, the eigenfrequencies are invariant $\Omega_n = \sqrt{A_nB_n}
\to \sqrt{\Lambda_n^2 A_n B_n/\Lambda_n^2} = \Omega_n$.
%


\subsection{Correlation functions}

Furthermore the observables should not depend on the particular normalization
of the basis functions.
The relative density-density correlations at time $t$ read
    \begin{alignat}{2}
	\frac{\langle \delta\hat\varrho(\bm x) \delta\hat\varrho(\bm x') \rangle}
		{\langle\hat\varrho(\bm x)\rangle\langle\hat\varrho(\bm x')\rangle}
	\, & = \, \frac{h_{n;t}^+(\bm x) h_{m;t}^+(\bm x')}
			{\sqrt{\varrho_0(\bm x)\varrho_0(\bm x')}}
		\langle \hat X_{n;t}^+ \hat X_{m;t}^+ \rangle
	\nn
	\, & = \, \frac{h_{n;t}^+(\bm x) h_{m;t}^+(\bm x')}
			{\sqrt{\varrho_0(\bm x)\varrho_0(\bm x')}}
		G_{n;t}^k \bar G_{m;t}^k
	\label{ddcorr}
    \end{alignat}
where I used the instantaneous eigenmode basis $h_{n;t}^+$, see Eq.\
\eqref{insteigenmode}.
Obviously, the factors $\Lambda_n$ and $1/\Lambda_n$ contributed by
$h_{n;t}^+$ and $\hat X_{n;t}$ cancel each other and the spatial correlations
are independent of the normalization $\Lambda_n$.
Similarly, the expression for the spatial phase-phase correlations
    \begin{alignat}{2}
	\langle\delta\hat\phi(\bm x)\delta\hat\phi(\bm x') \rangle
	\, & = \,
		\frac{h_{n;t}^-(\bm x) h_{m;t}^-(\bm x')}
			{\sqrt{\varrho_0(\bm x) \varrho_0(\bm x') }}
		\langle \hat X_{n;t}^- \hat X_{m;t}^- \rangle
	\nn
	\, & = \,
		\frac{h_{n;t}^-(\bm x) h_{m;t}^-(\bm x')}
			{\sqrt{\varrho_0(\bm x) \varrho_0(\bm x') }}
		F_{n;t}^k \bar F_{m;t}^k \,.
	\label{ppcorr}
    \end{alignat}
yields the same result regardless of the employed basis.

Hence, the factors $\Lambda_n$ can be chosen at will.
There exist, however, several convenient choices for 
$\Lambda_n$:
firstly, the density modes might be normalized to unity,
$\int d^Dx (h_n^+)^2 = 1$.
This is advantageous if left and right eigenvectors are the same,
i.e., if $\mathcal A\mathcal B$ is symmetric.
The drawback is that if $\mathcal A\mathcal B$ is not symmetric and
therefore $h_n^+ \neq h_n^-$, only one of the mode functions, $h_n^+$, can
be normalized to unity, whereas the norm of the $h_n^-$ follows from
Eq.\ \eqref{dual}.
And, secondly, these factors $\Lambda_n$ can be fixed by demanding
$A_n = B_n = \Omega_n$.
In this case, the prefactors of \eqref{ineigenmodes} initially obey
$F_n^m(t_{\rm in}) = \delta_{nm}/\sqrt{2}$ and
$G_n^m(t_{\rm in}) = i \delta_{nm}/\sqrt{2}$ and both quadratures
$\langle (\hat X_n^\pm)^2 \rangle(t_{\rm in}) = 1/2$.
However, one should note that neither of the eigenfunctions $h_n^\pm$ is
generally normalized to unity, $\int d^Dx (h_n^\pm)^2 \neq 1$.

Of course, the coefficients $\langle \hat X_{n;t}^+ \hat X_{m;t}^+\rangle$
and $\langle \hat X_{n;t}^- \hat X_{m;t}^- \rangle$ do depend on the
particular choice of the basis functions $h_{n;t}^\pm$ and thus also on
the factors $\Lambda_n$.
For instance, the adiabatic density-density and phase-phase correlations
of a particular mode read
    \begin{alignat}{2}
	\left\langle \hat X_{n;t}^- \hat X_{n;t}^- \right\rangle_{\rm ad}(t)
		\, & = \, \frac{A_{n;t}}{2\Omega_{n;t}}
		\, & \propto & \, \frac{1}{\Lambda_n^2}
	\,,\nn
	\left\langle \hat X_{n;t}^+ \hat X_{n;t}^+ \right\rangle_{\rm ad}(t)
		\, & = \, \frac{\Omega_{n;t}}{2A_{n;t}}
		\, & \propto & \, \Lambda_n^2 \,.
	\label{adiabcorr}
    \end{alignat}
The dependence on the factor $\Lambda_n$ becomes important regarding the
low excitations in the Thomas-Fermi approximation, cf.\ Sec.\ \ref{SECTF}:
the modes decouple and it is not necessary to introduce a
new spatial basis $h_n^\pm$ at the time of measurement.
One has instead
$A_{n;t} = f^2 A_{n;t_0}$ and $B_{n;t} = B_{n;t_0}$ such that
$\langle \hat X_{n;t}^+ \hat X_{n;t}^+ \rangle_{\rm ad} \propto f$
while
$\langle \hat X_{n;t}^- \hat X_{n;t}^- \rangle_{\rm ad} \propto 1/f$,
i.e., the phase and density fluctuations apparently increase or decrease
even for adiabatic evolution.
In view of this $\Lambda_n$ ambiguity, the (absolute) density-density
or phase-phase correlations provide no adequate measure for the squeezing
(i.e., non-adiabaticity) of a single mode.
The Fourier transforms of Eqs.\ \eqref{ddcorr} and \eqref{ppcorr} on the
other hand are independent of $\Lambda_n$ but do not represent the excitation
eigenmodes.


\subsection{Bogoliubov transformation and particle production}

As will be shown in the following, the (instantaneous) quasi-particle
number measures the relative deviation of density and phase fluctuations
from their adiabatic values.
In view of the different expansions \eqref{eigenmode} and
\eqref{insteigenmode} of density and phase fluctuations into their initial
and adiabatic eigenfunctions, the corresponding
creation and annihilation operators $\hat a_n^\dagger$, $\hat a_n$
and $\hat b_{n;t}^\dagger$, $\hat b_{n;t}$, respectively, can be transformed
by virtue of a Bogoliubov transformation.
For the annihilators $\hat b_{n;t}$ follows in particular
    \begin{alignat}{2}
	\hat b_{n;t}
		\, & = \,
			\sqrt{\frac{\Omega_{n;t} }{2 A_{n;t}}}
				\hat X_{n;t}^-
			- i \sqrt{\frac{A_{n;t}}{2\Omega_{n;t}}}
				\hat X_{n;t}^+
	\nn
		\, & = \, \alpha_{nm}(t) \hat a_m +
			\beta_{nm}(t) \hat a_m^\dagger \,
	\label{bogo}
    \end{alignat}
with the Bogoliubov coefficients $\alpha_{nm}$ and $\beta_{nm}$.
The first line follows from inversion of Eq.\ \eqref{instmodes}
and the second line can be inferred after transforming
\eqref{ineigenmodes} to the new basis $h_{n;t}^\pm$.
%
Since $\beta_{nm} \neq 0$ for non-adiabatic evolution,
the quasi-particle number operator
$\hat N_{n}(t) = \hat b_{n;t}^\dagger \hat b_{n;t}$
will have a non-zero expectation value as well
    \begin{alignat}{2}
	N_n(t) \, & = \,
		\left\langle \hat b_{n;t}^\dagger \hat b_{n;t} \right\rangle
		\, = \, \sum_m |\beta_{nm}(t)|^2 
	\nn
		\, & = \
		\frac{\Omega_{n;t}}{2A_{n;t}}
			\left\langle (\hat X_{n;t}^-)^2 \right\rangle
		+ \frac{A_{n;t}}{2\Omega_{n;t}}
			\left\langle (\hat X_{n;t}^+)^2 \right\rangle
	\nn
		&\qquad+ \frac i2 \left\langle
			\left[\hat X_{n;t}^+,\hat X_{n;t}^-\right]
		\right\rangle
    \end{alignat}
where the commutator $\left[\hat X_{n;t}^+, \hat X_{n;t}^-\right] = -i$.
Noting that the prefactors $\Omega_{n;t}/2A_{n;t}$ and
$A_{n;t}/2\Omega_{n;t}$ are just the adiabatic density-density and
phase-phase correlations, see Eq.\ \eqref{adiabcorr},
the particle number can be rewritten
    \begin{alignat}{2}
	N_n(t)
		\, & = \,
		\frac{ \langle(\hat X_{n;t}^-)^2\rangle
			- \frac{A_{n;t}}{2\Omega_{n;t}}}
				{2 A_{n;t}/\Omega_{n;t}}
		+
		\frac{ \langle(\hat X_{n;t}^+)^2\rangle
			- \frac{\Omega_{n;t}}{2A_{n;t}}}
				{2 \Omega_{n;t}/A_{n;t}} \,,
	\label{instpartnum}
    \end{alignat}
i.e., the instantaneous particle number gives just the relative deviation
of the density and phase correlations from their adiabatic values.
Note that expression \eqref{instpartnum} does not contain the correlations
between different modes.
To this end, it would be necessary to evaluate
$\langle \hat N_n \hat N_m \rangle$, which is fourth order in the
$\hat b_{n;t}$.
This observable can be reduced to expectation values quadratic
in in the $\hat b_{n;t}$ by virtue of Wick's theorem, see, e.g.,
\cite{Giorgini2}.


\subsection{Scaling}

Another interesting aspect regards the scaling of the correlation
functions \eqref{ddcorr} and \eqref{ppcorr} with the interaction
strength:
within the Thomas-Fermi approximation, see Sec.\ \ref{SECTF}, the
evolution equations \eqref{eigenmodeeomTF} are independent of $g_0$.
All properties of the linear excitations are determined by
the chemical potential $\mu_0$ and the variations $g(t)/g_0$ and
$\omega(t)/\omega_0$, in particular the expectation values
$\langle \hat X_{n;t}^\pm \hat X_{m;t}^\pm\rangle$.
Only the normalization factor
$\sqrt{\varrho_0(\bm x)\varrho_0(\bm x')} =
(1/g_0)\sqrt{[\mu - V(\bm x)][\mu_0 - V(\bm x')]}$
in Eqs.\ \eqref{ddcorr} and \eqref{ppcorr} depends on $g_0$.
Hence, the relative density-density and phase-phase correlations in
the Thomas-Fermi approximation are both proportional to $g_0$
for fixed $\mu_0$.


\begin{thebibliography}{9999}
\newfont{\cyr}{wncyr10}

\bibitem{Manybody}
I.\ Bloch, J.\ Dalibard, and W.\ Zwerger,
Rev.\ Mod.\ Phys.\ 80, 885 (2008).

\bibitem{BEC}
L.\ Pitaevskii and S. Stringari,
{\em Bose-Einstein Condensation}
(Oxford University Press, Oxford, UK, 2003);
A.\ J.\ Leggett,
Rev.\ Mod.\ Phys. {\bf 73}, 307 (2001).

\bibitem{Mott}
M.\ Greiner {\em et al.},
Nature {\bf 415}, 39 (2002);
F.\ Gerbier {\em et al.},
Phys.\ Rev.\ Lett.\ {\bf 96}, 090401 (2006);
S.\ F\"olling {\em et al.},
Phys.\ Rev.\ Lett.\ {\bf 97}, 060403 (2006).

\bibitem{BHM}
M.\ P.\ A.\ Fisher, P.\ B.\ Weichman, G.\ Grinstein, and D.\ S.\ Fisher,
Phys. Rev. B {\bf 40}, 546 (1989);
D.\ Jaksch {\em et al.},
Phys.\ Rev.\ Lett.\ {\bf 81}, 3108 (1998);
D.\ Jaksch and P.\ Zoller,
Ann.\ Phys.\ (N.Y.) {\bf 315}, 52 (2005);
R.\ Sch\"utzhold, M.\ Uhlmann, Y.\ Xu, and U.\ R.\ Fischer,
Phys.\ Rev.\ Lett.\ {\bf 97}, 200601 (2006);
C.\ Kollath, A.\ M.\ L\"auchli, and E.\ Altman,
Phys.\ Rev.\ Lett.\ {\bf 98}, 180601 (2007).

\bibitem{Volovik}
G.\ Volovik,
{\em The Universe in a Helium Droplet}
(Oxford University Press, Oxford, UK, 2003).

\bibitem{Review}
C.\ Barcel\'o, S.\ Liberati, and M.\ Visser,
Living Rev. Relativity\ {\bf 8}, 12 (2005).

\bibitem{Uwe2}
U.\ R.\ Fischer,
Mod.\ Phys.\ Lett.\ A {\bf 19}, 1789 (2004).

\bibitem{Piyush}
P.\ Jain, S.\ Weinfurtner, M.\ Visser, and C.\ W.\ Gardiner,
Phys.\ Rev.\ A {\bf 76}, 033616 (2007).

\bibitem{Silke}
S.\ Weinfurtner, A.\ White, and M.\ Visser,
Phys.\ Rev.\ D {\bf 76}, 124008 (2007).

\bibitem{Barcelo}
C.\ Barcel\'o, S.\ Liberati, and M.\ Visser,
Phys.\ Rev.\ A {\bf 68}, 053613 (2003);
%
C.\ Barcel\'o, S.\ Liberati, and M.\ Visser,
Int.\ J.\ Mod.\ Phys.\ D {\bf 12}, 1641 (2003);
C.\ Barcel\'o, S.\ Liberati, and M.\ Visser,
Class.\ Quant.\ Grav.\ {\bf 18}, 1137 (2001);
{\em ibid.} {\bf 18}, 3595 (2001).


\bibitem{Uwe}
U.\ R.\ Fischer and R.\ Sch\"utzhold,
Phys.\ Rev.\ A {\bf 70}, 063615 (2004).

\bibitem{ich_njp}
M.\ Uhlmann, Yan Xu, and R.\ Sch\"utzhold,
New J.\ Phys.\ {\bf 7}, 248 (2005).

\bibitem{PeterF}
P.\ O.\ Fedichev and U.\ R.\ Fischer,
Phys.\ Rev.\ A {\bf 69}, 033602 (2004).


\bibitem{Kurita}
Y.\ Kurita and T.\ Morinari,
Phys.\ Rev.\ A {\bf 76}, 053603 (2007).

\bibitem{Matt}
M.\ Visser and S.\ Weinfurtner,
Phys.\ Rev.\ D {\bf 72}, 044020 (2005).




\bibitem{Garay}
L. J. Garay, J. R. Anglin, J. I. Cirac, and P. Zoller,
Phys.\ Rev.\ Lett.\ {\bf 85}, 4643 (2000)

\bibitem{WuesterBlackhole}
S.\ W\"uster,
Phys. Rev. A {\bf 78}, 021601(R) (2008);
S.\ W\"uster and C.\ M.\ Savage,
Phys. Rev. A {\bf 76}, 013608 (2007).

\bibitem{Unruh}
W.\ G.\ Unruh,
Phys.\ Rev.\ Lett.\ {\bf 46}, 1351 (1981);
W.\ G.\ Unruh,
Phys.\ Rev.\ D {\bf 51}, 2827 (1995).

\bibitem{artificialblackholes}
M.\ Novello, M.\ Visser, and G.\ Volovik (eds.),
{\em Artificial Black Holes}
(World Scientific, Singapore, 2002).

\bibitem{Hawking}
S.\ W.\ Hawking,
Nature {\bf 248}, 30 (1974);
S.\ W.\ Hawking,
Commun.\ math.\ Phys.\ {\bf 43}, 199 (1975).

\bibitem{LL}
A.\ R.\ Liddle and D.\ H.\ Lyth,
{\em Cosmological Inflation and Large-Scale Structure},
(Cambridge University Press, Cambridge, UK, 2000).

\bibitem{BirrellDavis}
N.\ D.\ Birrell and P.\ C.\ Q.\ Davies,
{\em Quantum Fields in Curved Space}
(Cambridge University Press, Cambridge, UK, 1982).


\bibitem{noise-correlations}
%
S.\ F\"olling, {\em et al.},
Nature {\bf 434}, 481 (2005).

\bibitem{interference}
D. Hellweg {\em et al.},
Phys.\ Rev.\ Lett.\ {\bf 91}, 010406 (2003);
S.\ Hofferberth {\em et al.},
Nature {\bf 449}, 324 (2007);
Nature Phys.\ {\bf 4}, 489 (2008).


\bibitem{Bragg}
S.\ B.\ Papp {\em et al.},
preprint {\sf arXiv:0805.0295}.

\bibitem{ultracoldHBT}
M.\ Schellekens {\em et al.},
Science {\bf 310}, 648 (2005).

\bibitem{numberfluct}
J.\ Esteve {\em et al.},
Phys.\ Rev.\ Lett.\ {\bf 96}, 130403 (2006).

%



\bibitem{Griffin}
A.\ Griffin,
Phys.\ Rev.\ B {\bf 53}, 9341 (1996);
%
S.\ Giorgini
Phys.\ Rev.\ A {\bf 61}, 063615 (2000).

\bibitem{projectedGP}
M.\ J.\ Davis, S.\ A.\ Morgan, and K.\ Burnett,
Phys.\ Rev.\ Lett.\ {\bf 87}, 160402 (2001);
P.\ B.\ Blakie and M.\ J. Davis,
Phys.\ Rev.\ A {\bf 72}, 063608 (2005);
%
P.\ B.\ Blakie {\em et al.},
preprint {\sf arXiv:0809.1487}.

\bibitem{WuesterBosenova}
S.\ W\"uster, J.\ J.\ Hope, and C.\ M.\ Savage,
Phys.\ Rev.\ A {\bf 71}, 033604 (2005);
S.\ W\"uster {\em et al.},
Phys.\ Rev.\ A {\bf 75}, 043611 (2007).




\bibitem{dimless}
Dimensionless units can be introduced through the initial trap
frequency $\omega_0$ and the atom mass $m$.
%
All energies are then measured in units of $\omega_0$,
the dimensionless time is defined through $t = \omega_0 \tau$,
and the characteristic length $a_0 = 1/\sqrt{m\omega_0}$,
where $\hbar = 1$.

\bibitem{feshbach}
E.\ Tiesinga, B.\ J.\ Verhaar, and H.\ T.\ C.\ Stoof
Phys.\ Rev.\ A {\bf 47}, 4114 (1993);
S.\ Inouye {\em et al.},
Nature {\bf 392}, 151 (1998);
Ph.\ Courteille, R.\ S.\ Freeland, and D.\ J.\ Heinzen,
Phys. Rev. Lett. {\bf 81}, 69 (1998).

\bibitem{scaling}
Y.\ Castin and R.\ Dum,
Phys.\ Rev.\ Lett.\ {\bf 77}, 5315 (1996);
Yu.\ Kagan, E.\ L.\ Surkov, and G.\ V.\ Shlyapnikov,
Phys.\ Rev.\ A {\bf 54}, R1753 (1996).


\bibitem{meanfield}
M.\ Girardeau and R.\ Arnowitt,
Phys.\ Rev.\ {\bf 113}, 755(1959);
%
C.\ W.\ Gardiner,
Phys.\ Rev.\ A {\bf 56}, 1414, (1997);
%
M.\ D.\ Girardeau,
{\em ibid} {\bf 58}, 775 (1998).



\bibitem{GP}
E.\ P.\ Gross,
Nuovo Cimento {\bf 20}, 454 (1961);
L.\ P.\ Pitaevskii, Zh.\ Eksp.\ Teor.\ Fiz.\ {\bf 40}, 646 (1961)
[Sov.\ Phys.\ JETP {\bf 13}, 451 (1961)].

\bibitem{Bogoliubov}
N.\ N.\ Bogoliubov,
J.\ Phys.\ (Moscow), {\bf 11}, 23, (1947);
P.\ G.\ de Gennes,
{\em Superconductivity of Metals and Alloys}
(W.\ A.\ Benjamin, New York, 1966).


\bibitem{inhom}
As similar expansion of the linear fluctuations is also considered in
\cite{Fetter}.
%
There, the authors expand the linear field operator $\hat \chi$ of a
static/stationary condensate into eigenmodes and not the phase and
density fluctuations $\hat \chi_\pm$.
%
This leads to complex mode functions $v_n$ and $u_n$, while $h_n^\pm$
are real.

\bibitem{Fetter}
A.\ L.\ Fetter,
Ann.\ Phys.\ (N.Y.) {\bf 70}, 67 (1972);
M.\ Lewenstein and L.\ You,
Phys.\ Rev.\ Lett.\ {\bf 77}, 3489 (1996);
M.\ Naraschewski and R.\ J.\ Glauber,
Phys.\ Rev.\ A {\bf 59}, 4595 (1999).

\bibitem{Ralf}
R.\ Sch\"utzhold,
Phys.\ Rev.\ Lett.\ {\bf 97}, 190405 (2006).


\bibitem{densityandphasemodefunctions}
The eigenfunctions of density and phase fluctuations $\delta\hat\varrho$
and $\delta\hat\phi$ are defined slightly differently than those of
$\hat \chi_\pm$
%
    \begin{alignat}{3}
	\hat \chi_+ \, & = \, h_n^+ \hat X_n^+
		\, & = & \, \frac{h_n^\varrho}{\sqrt{\varrho_0}}
			\delta\hat\varrho_n
		\, & = & \, \frac{\delta\hat\varrho}{\sqrt{\varrho_0}}
	\nn
	\hat \chi_- \, & = \, h_n^- \hat X_n^-
		\, & = & \, \sqrt{\varrho_0} h_n^\phi \delta\hat\phi
		\, & = & \, \sqrt{\varrho_0}\delta\hat\phi \,.
	\nonumber
    \end{alignat}
Noting that $\na_x\varrho_0 \approx 0$ the center of the trap,
the eigenfunctions $h_n^+$ and $h_n^\varrho$ (and similarly $h_n^-$
and $h_n^\phi$) must be approximately proportional for low excitations.
%
Hence, the $X_n^\pm$ for $\Omega_n \ll \mu_0$ indeed represent the
proper density and phase eigenmodes $\delta\hat\varrho_n$ and
$\delta\hat\phi_n$ up to a constant prefactor $\varrho_0^{\pm 1/2}$.


\bibitem{Wald}
R.\ M.\ Wald, {\em General Relativity}
(University of Chicago Press, Chicago, IL, 1984).

\bibitem{MTW}
C.\ W.\ Misner, K.\ S.\ Thorne, and J.\ A.\ Wheeler,
{\em Gravitation}
(W.H. Freeman, San Francisco, USA, 1973).

\bibitem{Visser}
M.\ Visser
{\em Lorentzian Wormholes: From Einstein to Hawking}
(Springer, New York, 1996).

\bibitem{HawkingEllis}
S.\ W.\ Hawking and G.\ F.\ R.\ Ellis,
{\em The Large Scale Structure of Spacetime}
(Cambridge University Press, Cambridge, UK, 1973).

\bibitem{MattHor}
M.\ Visser,
Class.\ Quant.\ Grav.\ {\bf 15}, 1767 (1998).

\bibitem{RalfHor}
R.\ Sch\"utzhold,
Lect.\ Notes Phys.\ {\bf 718}, 5 (2007);
Class.\ Quant.\ Grav.\ {\bf 25}, 114011 (2008).

\bibitem{GP-modeexp}
C.\ M.\ Dion and E.\ Canc\`es,
Phys.\ Rev.\ E {\bf 67}, 046706 (2003).

\bibitem{Abramowitz}
{\em Handbook of Mathematical Functions}, edited by
M.\ Abramowitz and I.\ A.\ Stegun,
(Dover, New York, 1970).

%
%
%

\bibitem{Sachdev}
S.\ Sachdev,
{\em Quantum Phase Transitions}
(Cambridge University Press, Cambridge, UK, 2000).

\bibitem{spinor}
L.\ E.\ Sadler {\em et al.},
Nature {\bf 443}, 312 (2006);
%
H.\ Saito and M.\ Ueda,
Phys.\ Rev.\ A {\bf 72}, 023610 (2005);
%
H.\ Saito, Y.\ Kawaguchi, and M.\ Ueda,
Phys.\ Rev.\ Lett.\ {\bf 96}, 065302 (2006);
%
A.\ Lamacraft
Phys.\ Rev.\ Lett.\ {\bf 98}, 160404 (2007);
%
M.\ Uhlmann, R.\ Sch\"utzhold, and U.\ R.\ Fischer,
Phys.\ Rev.\ Lett.\ {\bf 99}, 120407 (2007).

\bibitem{Giorgini2}
S.\ Giorgini, L.\ P.\ Pitaevskii, and S.\ Stringari,
Phys.\ Rev.\ Lett.\ {\bf 80}, 5040 (1998).





















\end{thebibliography}
\end{document}